\documentclass[12pt]{iopart}
\usepackage{iopams}  
\usepackage[breaklinks=true,colorlinks=true,linkcolor=blue,urlcolor=blue,citecolor=blue]{hyperref}

\usepackage{graphicx}
\usepackage[space]{grffile}
\usepackage{latexsym}

\expandafter\let\csname equation*\endcsname\relax
\expandafter\let\csname endequation*\endcsname\relax

\usepackage{amsfonts,amsmath,amssymb}
\usepackage{url}
\usepackage[utf8]{inputenc}
\usepackage{hyperref}
\hypersetup{colorlinks=false,pdfborder={0 0 0},}
\usepackage{textcomp}
\usepackage{longtable}
\usepackage{multirow,booktabs}

\usepackage[T1]{fontenc}
\usepackage[utf8]{inputenc}
\usepackage{textcomp}
\usepackage{graphicx}
\usepackage{ae}
\usepackage{float} 
\usepackage{amsfonts}

\begin{document}

\title{Contactless Remote Induction of Shear Waves in Soft Tissues Using a Transcranial Magnetic Stimulation Device}

\author{Pol Grasland-Mongrain(1), Erika Miller-Jolicoeur(1), An Tang(2), Stefan Catheline(3,4), Guy Cloutier(1,5,6) }
\address{(1) Laboratoire de Biorh\'eologie et d’Ultrasonographie M\'edicale, Research Center of the Montreal University Health Centre, Montreal (QC), Canada}
\address{(2) Research Center of the Montreal University Health Centre, Montreal (QC), Canada}
\address{(3) Laboratoire de Th\'erapie et Applications des Ultrasons, Inserm u1032, Inserm, Lyon, F-69003, France}
\address{(4) Universit\'e Lyon 1 Claude Bernard, Lyon, F-69003, France}
\address{(5) D\'epartement de Radiologie, Radiooncologie et M\'edecine Nucl\'eaire, University of Montreal, Montreal (QC), Canada}
\address{(6) Institut de G\'enie Biom\'edical, Montreal (QC), Canada}

\ead{contact: pol.grasland-mongrain@ens-cachan.org, guy.cloutier@umontreal.ca}

\begin{abstract}
This study presents the first observation of shear wave induced remotely within soft tissues. It was performed through the combination of a transcranial magnetic stimulation device and a permanent magnet. A physical model based on Maxwell and Navier equations was developed. Experiments were performed on a cryogel phantom and a chicken breast sample. Using an ultrafast ultrasound scanner, shear waves of respective amplitude of 5 and 0.5 micrometers were observed. Experimental and numerical results were in good agreement. This study constitutes the framework of an alternative shear wave elastography method.%
\end{abstract}

\maketitle

\section{Introduction}

Propagation of elastic waves in solids has been described in various fields of physics, including geophysics, soft matter physics or acoustics. Elastic waves can be separated in two components in a bulk: compression waves, corresponding to a curl-free propagation; and shear waves, corresponding to a divergence-free propagation. Shear waves have drawn a strong interest in medical imaging with the development of shear wave elastography methods \cite{muthupillai1995magnetic}, \cite{sarvazyan1998shear}. These methods use shear waves to measure or map the elastic properties of biological tissues. Shear wave speed measurement permits calculation of the tissue shear modulus. Shear wave elastography techniques have been successfully applied to several organs such as the liver \cite{sandrin2003transient}, the breast \cite{berg2012shear}, the arteries \cite{schmitt2010ultrasound} and the prostate \cite{cochlin2002elastography}, to name a few examples. The brain has also been studied, and its elasticity is of strong interest for clinicians \cite{mariappan2010magnetic}, \cite{kruse2008magnetic}. For example, it has been shown that Alzeihmer's disease, hydrocephalus or multiple sclerosis are associated with changes in brain elastic properties \cite{murphy2011decreased}, \cite{taylor2004reassessment}, \cite{wuerfel2010mr}.

Clinical shear wave elastography techniques currently rely on an external vibrator \cite{muthupillai1995magnetic}, \cite{sandrin2003transient} or on a focused acoustic wave \cite{nightingale2002acoustic}, \cite{sarvazyan1998shear} as the shear wave source. However, these techniques are limited in situations where the organ of interest is located behind a strongly attenuating medium like the brain behind the skull and surrounded by the cerebrospinal fluid. While external shakers are able to transmit some shear waves, using acoustic, pneumatic, piezoelectric or electromagnetic actuators \cite{kruse2008magnetic}, \cite{20833495}, \cite{11548931}, \cite{Braun_2003}, this approach can be uncomfortable for patients. Alternatively, acoustic waves may be transmitted through the skull to induce shear waves inside the brain, but the skull attenuates and deforms the acoustic beam, preventing efficient transmission of energy. Recently, it has also been shown that physiological body motion can be used, via blood pulsation \cite{23008140}, \cite{Weaver_2012} or noise correlation \cite{gallot2011passive}, \cite{Zorgani_2015}, but these methods still require further development before clinical application in the context of brain elastography.

Recently, it was demonstrated that the combination of an electrical current and a magnetic field could create displacements which propagate as shear waves in biological tissues \cite{basford2005lorentz}, \cite{grasland2014elastoEMarticle}. If the electrical current is induced using a coil, this would allow the technique to remotely induce shear waves. In the case of brain elastography, this would allow inducing shear waves directly inside the brain.

To achieve this objective, we propose to use a transcranial magnetic stimulation (TMS) device \cite{hallett2000transcranial}. This instrument is used to induce an electrical current directly inside the brain by using an external coil. TMS is currently employed by neurologists to study brain functionality \cite{ilmoniemi1999transcranial} and by psychiatrists to treat depression \cite{sakkas2006repetitive}. TMS is occasionally combined with magnetic resonance imaging (MRI) \cite{devlin2003semantic}, \cite{bohning1997mapping}; however, no study has yet reported the production of shear waves when combining TMS and magnetic fields.

This article first presents the physical model describing the generation of shear waves resulting from the combination of a remotely induced electrical current and a magnetic field. It describes experiments performed in poyvinyl alcohol cryogel and biological tissue samples. A numerical study of the experiments is then presented. Results section shows a good consistency between experimental and numerical displacement maps. Some critical excitation parameters were investigated as well as dependence of the shear wave amplitude with the magnetic field and electrical current intensity. Practical implementation in a context of shear wave elastography of the brain is finally discussed.

  \section{Physical model}

\begin{figure}[h!]
\begin{center}
\includegraphics[width=0.98\columnwidth]{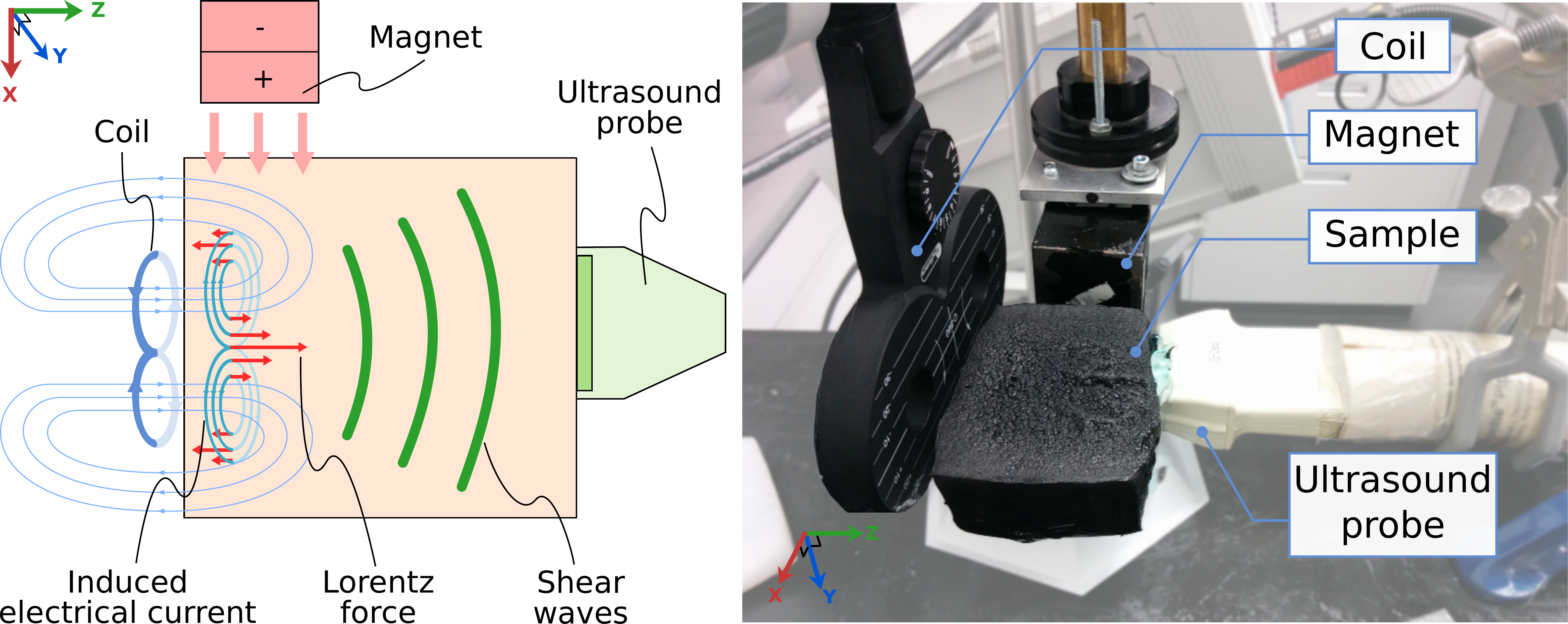}
\caption{\label{Figure1}(A) Scheme of the experiment. A coil is inducing remotely an electrical current (blue circles) in a sample. A magnet creates a magnetic field (pink arrows) in the sample. Combination of the electrical current and the magnetic field induces a Lorentz force (red arrows). This force creates displacements which propagate as shear waves (green waves) tracked through an ultrasound probe. (B) Experimental setup. The tested sample is a polyvinyl alcohol tissue-mimicking phantom (plastic box surrounding the phantom was removed for clarity). The electrical current was applied by the coil. The magnetic field was created by the magnet. Ultrasound images were acquired through the probe coupled to the sample.%
}
\end{center}
\end{figure}

We set up the experiment illustrated in Figure \ref{Figure1}-(A). The key components are as follows: a coil induces an electrical current $\mathbf{j}$ in the sample; a magnet creates a magnetic field $\mathbf{B}$; an ultrasound probe tracks displacements $\mathbf{u}$ propagating as shear waves in the sample. X is defined as the main magnetic field axis, Z as the main ultrasound propagation axis, and Y an axis orthogonal to X and Z following the right-hand rule. The origin of coordinates (0,0,0) is located in the middle of the coil (i.e., between the two loops).

For a circular coil centered in (0,0,0) of linear element $d\mathbf{l}$ crossed by an electrical current $I(t)$, using Coulomb gauge (i.e., $\nabla . \mathbf{A} = 0$ where $\mathbf{A}$ is the magnetic potential vector), and negligible propagation time of electromagnetic waves, the electrical field $\mathbf{E}(\mathbf{r},t)$ along space $\mathbf{r}$ and time $t$ is equal to \cite{jackson1998classical}:
\begin{equation}
\mathbf{\mathbf{E(\mathbf{r},t)}} = - \nabla \Phi - \frac{d I}{d t} \frac{N \mu_0}{4\pi}\int{\frac{\mathbf{dl}}{r}}
\label{Equation1}
\end{equation}
where $\Phi$ is the electrostatic scalar potential, $N$ is the number of turns of the coil and $\mu_0$ is the magnetic permeability of the coil material. In an unbounded medium, $\Phi$ is only due to free charges \cite{grandori1991magnetic}, that we supposed negligible in our case. Being additive, the total electrical field created by two or more coils is simply the sum of the contribution of each coil. The induced electrical current density $\mathbf{j}$ is retrieved using the local Ohm's law $\mathbf{j}=\sigma \mathbf{E}$, where $\sigma$ is the electrical conductivity of the medium.

The body Lorentz force $\mathbf{f}$ can then be calculated using the relationship $\mathbf{f} = \mathbf{j} \times \mathbf{B}$, where $\mathbf{B}$ is the magnetic field created by the permanent magnet. Considering the tissue as an elastic, linear and isotropic solid, Navier's equation governs the displacement $\mathbf{u}$ at each point of the tissue submitted to an external body force $\mathbf{f}$ \cite{aki1980quantitative}:
\begin{equation}
\rho\frac{d^2\mathbf{u}}{dt^2} = (K + \frac{4}{3}\mu) \nabla (\nabla . \mathbf{u}) + \mu \nabla \times (\nabla \times \mathbf{u}) + \mathbf{f}
\label{Equation3}
\end{equation}
where $\rho$ is the medium density, $\mathbf{u}$ the local displacement, $K$ the bulk modulus and $\mu$ the shear modulus.

Using Helmholtz decomposition $\mathbf{u}=\nabla \phi + \nabla \times \mathbf{\psi}$, where $\phi$ and $\mathbf{\psi}$ are respectively a scalar and a vector field, two elastic waves can be retrieved: a compression wave, propagating at a celerity $c_k = \sqrt{(K+\frac{4}{3}\mu)/\rho}$, and a shear
wave, propagating at celerity  $c_s = \sqrt{\mu/\rho}$ \cite{sarvazyan1998shear}. As $\rho$ varies typically by a few percent between different soft tissues \cite{cobbold2007foundations}, we can suppose an homogeneous density, and measuring $c_s$ allows to compute the shear modulus $\mu$ of the tissue.

\section{Material and methods}

\begin{figure}[h!]
\begin{center}
\includegraphics[width=0.98\columnwidth]{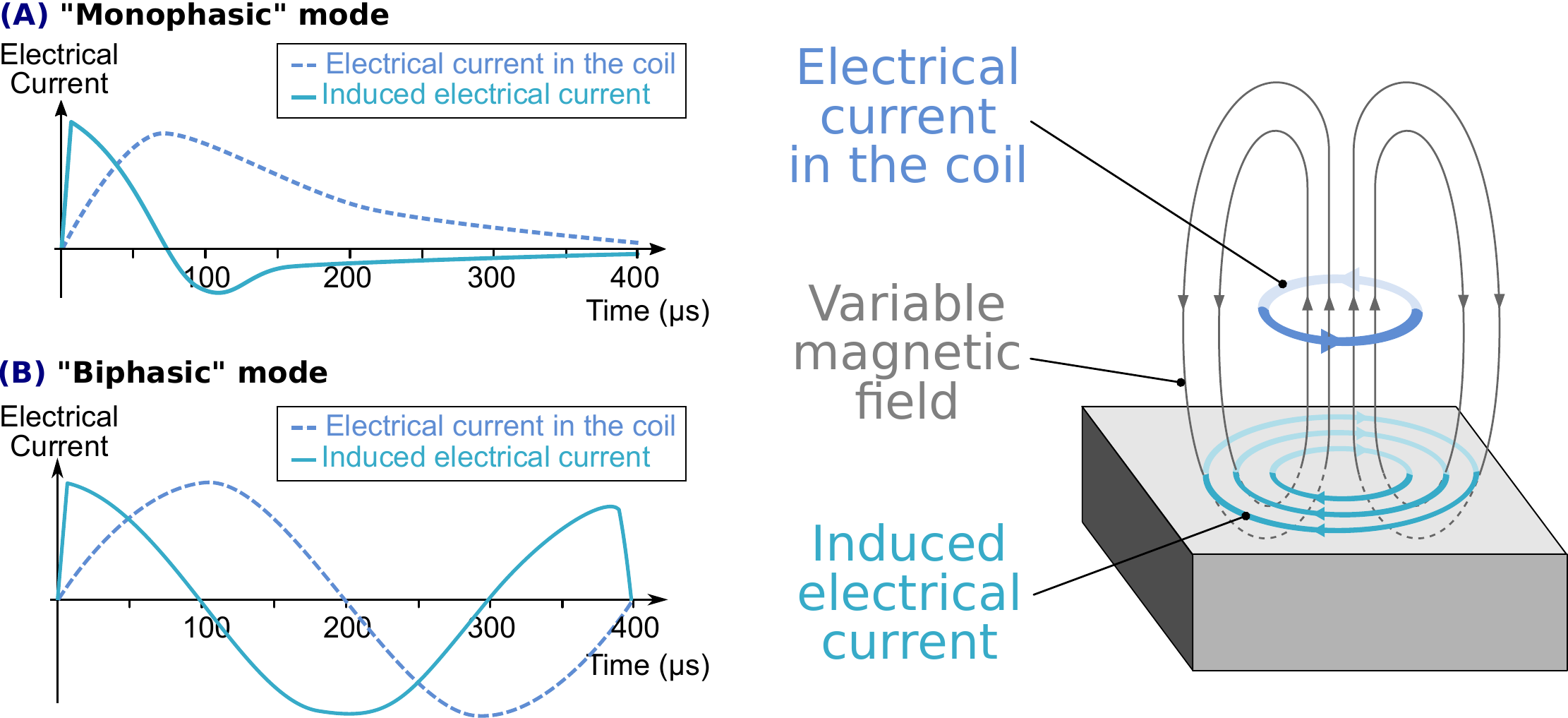}
\caption{\label{InducedElectricalCurrent} (A) Electrical current in the coil and induced electrical current when used in "monophasic" mode. (B) Electrical current in the coil and induced electrical current when used in "biphasic" mode. (Right) Scheme of the induction of electrical current in the medium by the TMS coil.%
}
\end{center}
\end{figure}

In the experimental setup, pictured in Figure \ref{Figure1}-(B), the electrical current was induced by a clinical TMS device using a 2x75 mm diameter coil (MagPro R100 device with C-B60 Butterfly coil, MagVenture, Farum, Danemark). The coil was placed 1 cm away from the medium, without any contact, and fixed to an independent support. The electrical current in the coil was in "monophasic" mode, i.e., a half cycle of 0.4 ms with a rising time of 70 $\mu$s, as illustrated in Figure \ref{InducedElectricalCurrent}-(A). Alternatively, "biphasic" mode, i.e., a full sinus cycle of 0.4 ms could be used, as illustrated in Figure \ref{InducedElectricalCurrent}-(B). According to the manufacturer's specifications, at 100\% amplitude, current reached a magnitude of 149.10$^6$ A.s$^{-1}$ in the coil (i.e., 30 kT.s$^{-1}$ during rising time), leading to a peak transient magnetic field of 2 T.s$^{-1}$ at the surface of the coil and of 0.74 T$^{-1}$ (i.e., 12 kT.s$^{-1}$ during rising time) at 20 mm in depth.

The magnetic field was induced by a 5x5x5 cm$^3$ N48 NdFeB magnet (model BY0Y0Y0, K\&J Magnetics, Pipersville, PA, USA). The magnet was placed 1 cm away from the medium, without any contact, and fixed to a second independent support. In the medium location, the magnetic field intensity ranged from 100 to 200 mT, as measured by a gaussmeter (Model GM2, AlphaLab, Salt Lake City, UT, USA).

Main tested sample was a 4x8x8 cm$^3$ water-based tissue-mimicking phantom made with 5\% polyvinyl alcohol (PVA), 0.1\% graphite powder and 5\% NaCl, giving a theoretical electrical conductivity of 7.5 S.m$^{-1}$. Three freezing/thawing cycles were applied to stiffen the material \cite{fromageau2007estimation}. The graphite powder (\#282863 product, Sigma-Aldrich,  Saint-Louis, MO, USA) was made of submillimeter particles, which presented a speckle pattern on ultrasound images. The sample was placed in a rigid plastic box of 2 mm thick layers with an opening on a side to introduce the ultrasound probe. The rigid box simulated a solid interface such as a skull and ensure also that any observed movement was not due to surrounding displacement of air. Alternatively, we used a similar phantom made of 5\% PVA, 0.1\% graphite powder and 2\% NaCl, giving a theoretical electrical conductivity of 3.5 S.m$^{-1}$. A biological tissue sample was also tested. This tissue was a chicken breast sample bought in local grocery of approximately 3x5x5 cm$^3$. It was degassed in a 20$^o$C saline water (0.9\% NaCl) during two hours prior to the experiment.

Each sample was observed with a 5 MHz ultrasonic probe made of 128 elements (ATL L7-4, Philips, Amsterdam, Netherlands) coupled to a Verasonics scanner (Verasonics V-1, Redmond, WA, USA). The probe was in contact with the sample with an ultrasound coupling gel but was fixed on a third independent support. It was used in ultrafast mode \cite{bercoff2004supersonic}, to acquire 1000 frames per second using plane waves and Stolt's fk migration algorithm \cite{garcia2013stolt}. The Z component of the displacement in the sample was observed by performing cross-correlations between radiofrequency images with a speckle-tracking technique, using 128x5 pixels$^2$ cross-correlation windows \cite{montagnon2012real}. Noise was partly reduced using a low-pass frequency filter (cut-off frequency at 1 kHz). Time $t$ = 0 ms was defined as the electrical burst emission.

Great care was taken to ensure that the three supports were not in contact and fixed separately. It could ensure that any vibration of one of the element could not be transmitted to the medium.

\section{Numerical study}
Additionally to the experiments, a three dimensional simulation of the experiments was performed using Matlab (Matlab 2010, The MathWorks, Natick, MA, USA). The numerical study was performed by (1) calculating the electrical current induced by the coil, (2) simulating the magnetic field created by the permanent magnet, (3) computing the resulting Lorentz force inside the medium, and finally (4) computing the propagation along space and time of the displacement due to the Lorentz force.

\begin{figure}[h!]
\begin{center}
\includegraphics[width=1\columnwidth]{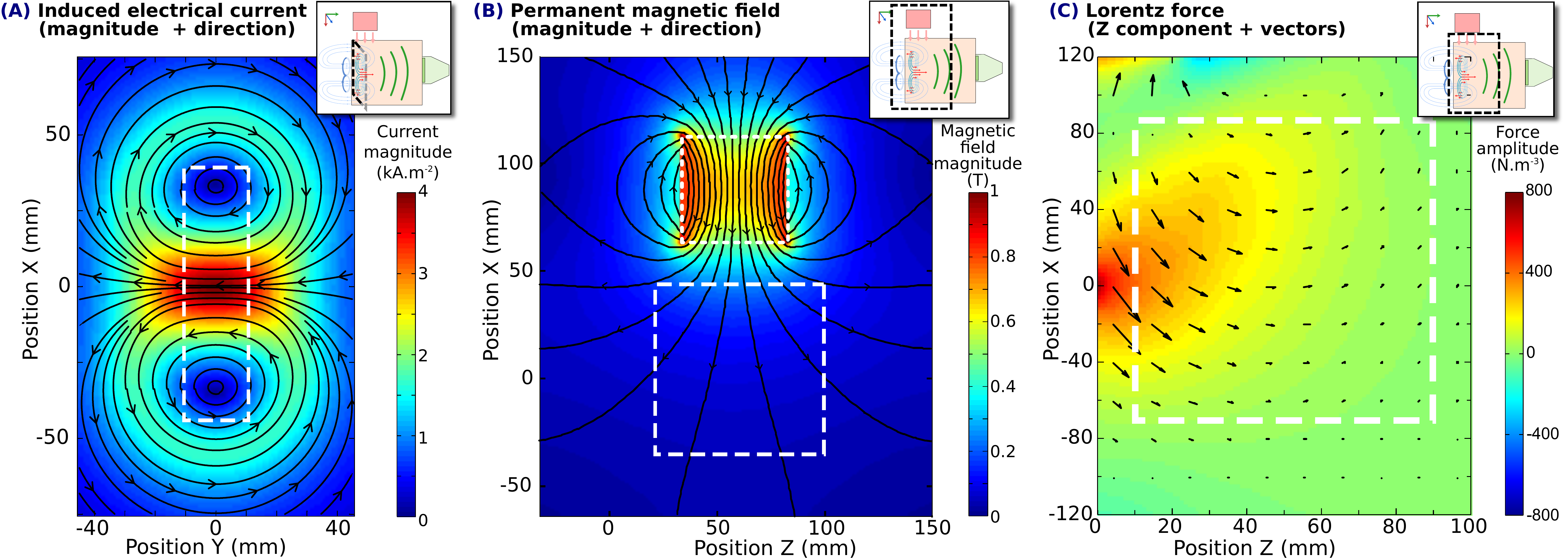}
\caption{\label{Figure3}(A) Electrical current induced by two 75 mm-diameter coils, in a XY plane in a 7.5 S.m$^{-1}$ medium at 2 cm of the coil, as calculated analytically. Black lines are representing the electrical current lines and colors the magnitude. Electrical current reached a magnitude of 4 kA.m$^{-2}$ in the medium location (dashed line). (B) Magnetic field as simulated by the Finite Element Magnetic Method software from a 5x5 cm$^2$ NdFeB magnet (dotted line). The magnetic field ranged from 100 to 200 mT at the medium location (dashed line). Black lines are representing the magnetic field lines and colors the magnitude. (C) Lorentz force in a XZ plane in the medium, as calculated from electrical current and magnetic field. Arrows are representing force vectors and colors the amplitude of the Z component. Lorentz force reached a magnitude of 600 N.m$^{-3}$ in the medium location (dashed line).%
}
\end{center}
\end{figure}

Using Equation \ref{Equation1} with two 75 mm diameter coils crossed by a 149.10$^6$ A.s$^{-1}$ electrical current, representing the TMS coil used in the experiment, electrical field $E$ was calculated in a 20x10x20 cm$^3$ volume (see \cite{Grandori_1991} for details on mathematical solving). Using Ohm's law, the electrical current $\mathbf{j}$ was estimated assuming an electrical conductivity $\sigma$ = 7.5 S.m$^{-1}$. No border effect has been taken into account. Induced electrical current in a XY plane at a depth of 2 cm with 2x2 mm$^2$ pixels is illustrated in Figure \ref{Figure3}-(A), with colors indicating the absolute magnitude and arrows the direction. The electrical current reached a density of 4 kA.m$^{-2}$ at the medium location.

A finite element software (Finite Element Magnetic Method \cite{FEMM}) was used to produce a two dimensional simulation of the magnetic field $\mathbf{B}$. The magnetic field was supposed to be approximately constant in the sample along the Y axis. The magnetostatic problem was solved from equations $\nabla \times H = \nabla \times M$, $\nabla B = 0$ and $B=\mu_p H$, with $H$ magnetic field intensity, $M$ magnetization of the medium, and $\mu_p$ medium permeability. Medium was considered as linear, and space was meshed with approximately 0.5 cm$^2$ triangles. The software simulated a N48 NdFeB permanent magnet of 5x5 cm$^2$ placed in a 30x30 cm$^2$ surface of air. Resulting magnetic field in a XZ plane is illustrated in Figure \ref{Figure3}-(B), with colors indicating the absolute magnitude and arrows the direction. The magnetic field ranged from 100 to 200 mT at the medium location.

The body Lorentz force $\mathbf{f}$ was computed from the cross-product of $\mathbf{j}$ and $\mathbf{B}$. The resulting Lorentz force in a XZ plane with 2x2 mm$^2$ pixels is illustrated in Figure \ref{Figure3}-C, with arrows indicating the Lorentz force vector and color its amplitude along Z - as the electrical current is induced in the XY plane and the magnetic field essentially along X direction, Lorentz force is mainly along Z direction. Lorentz force reached a magnitude of 600 N.m$^{-3}$ in the medium location.

Finally, displacement $\mathbf{u}(\mathbf{r},t)$ was determined analytically along space (pixels of 2x2 mm$^2$) and time (steps of 1 ms) by solving Equation \ref{Equation3} with the Green operator \cite{aki1980quantitative}. It used a medium density $\rho$ of 1000 kg.m$^{-3}$, a bulk modulus $K$ of 2.3 GPa and a shear modulus $\mu$ of 16 kPa, corresponding to a shear wave speed of 4 m.s$^{-1}$.

\section{Results}
\begin{figure}[h!]
\begin{center}
\includegraphics[width=1\columnwidth]{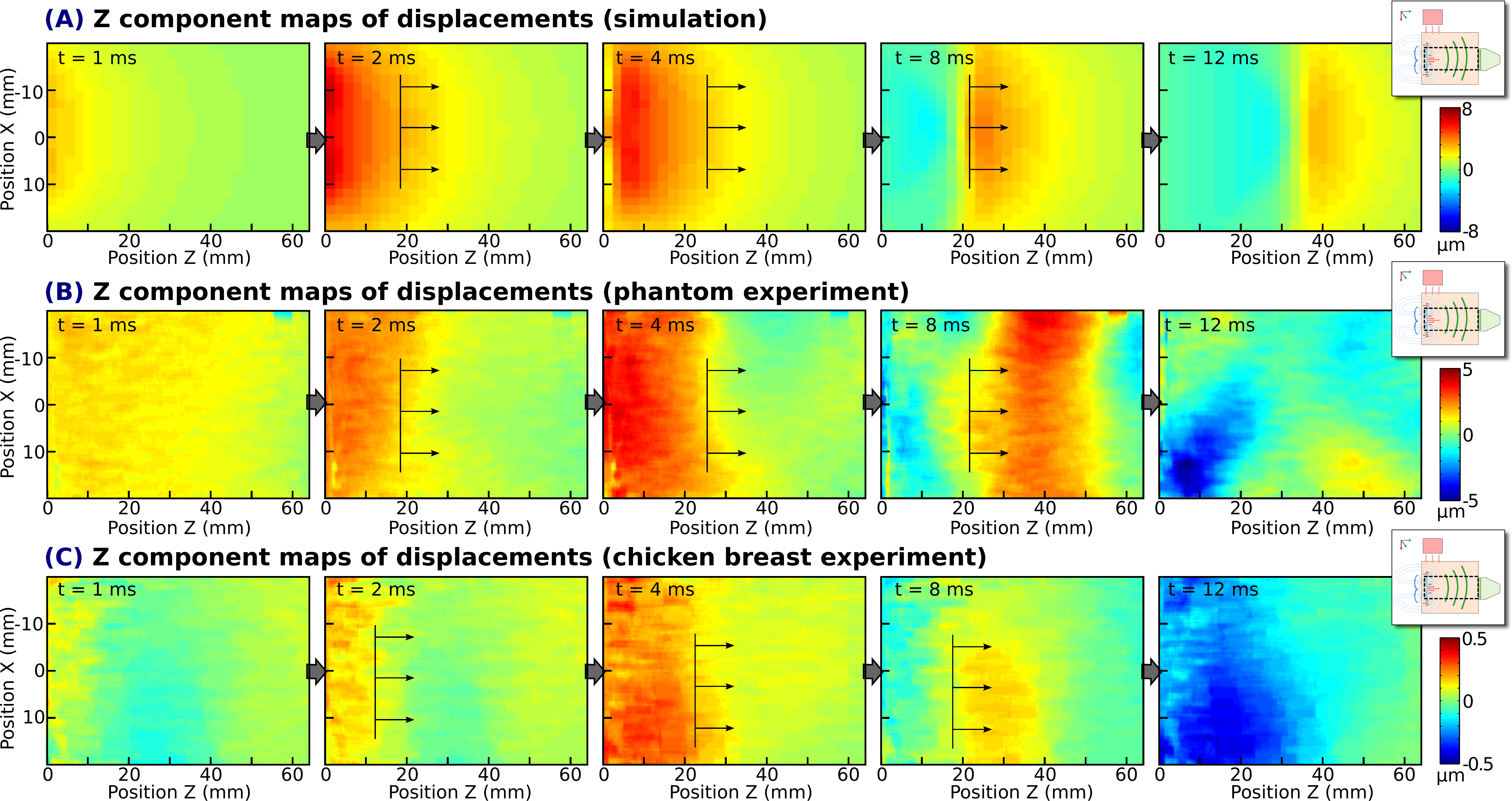}
\caption{\label{Figure4}Z component maps of the displacement over time, respectively 1, 2, 4, 8 and 12 ms after current emission, as given by the simulation (A), the experiment on the PVA phantom (B) and on the chicken sample (C). A shear wave can be observed in the three cases, with hand drawn black arrows indicating propagation of the wave front.%
}
\end{center}
\end{figure}

Z component maps of the displacements over time are illustrated in Figure \ref{Figure4}, respectively 1, 2, 4, 8 and 12 ms after current emission, as given by the simulation (A), experiment in the PVA phantom (B) and experiment in the chicken breast sample (C). Initial displacements occurs where the Lorentz force has the highest magnitude, on the opposite side of the ultrasound probe, so displacement is not due to probe vibration. Displacements reached an amplitude of 5 $\mu$m in the phantom and 0.5 $\mu$m in the chicken sample. Displacement maps were harder to compute in the chicken breast sample, as electrical conductivity was lower and as speckle was of poorer quality. They propagated as shear waves, whose speed was 4 m.s$^{-1}$ for the simulation, 4.0$\pm$1.0 m.s$^{-1}$ for the PVA phantom and 3.5$\pm$1.0 m.s$^{-1}$ for the chicken sample along Z axis. These values correspond respectively to a Young's modulus of 48$\pm$24 kPa for the PVA phantom and of 37$\pm$20 kPa for the chicken sample.

\begin{figure}[h!]
\begin{center}
\includegraphics[width=1\columnwidth]{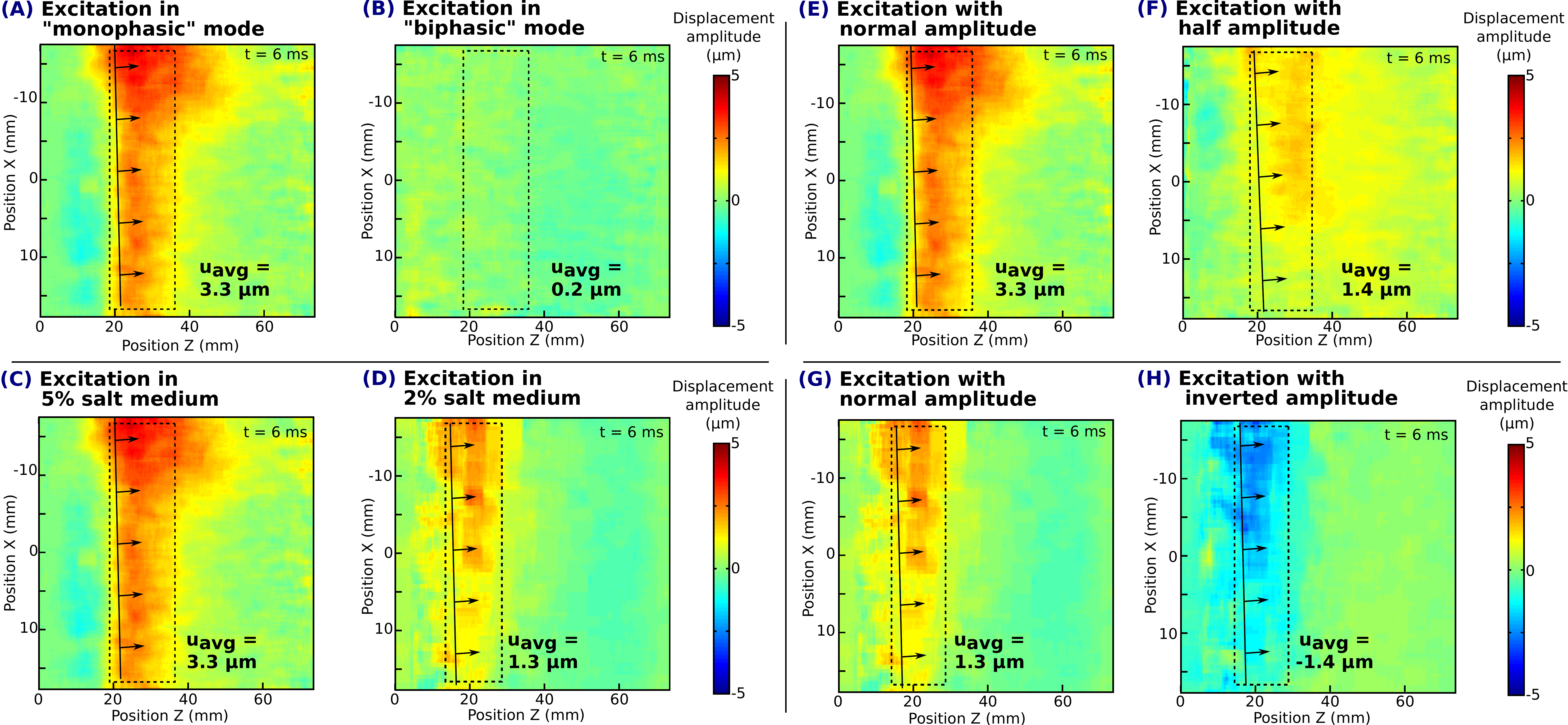}
\caption{\label{Comparaison} (A-B) Z component map with excitation with "monophasic" and "biphasic" mode respectively. Shear wave can be observed in "monophasic" mode, but no displacement occurs in "biphasic" mode. (C-D) Z component map with excitation in a 5\% salt medium and a 2\% salt medium respectively. Amplitude of displacement is approximately divided by the same factor as electrical conductivity. (E-F) Z component map with excitation with 100\% amplitude and 50\% amplitude respectively. Amplitude of displacements is roughly divided by two when amplitude of excitation is halved. (G-H) Z component map with excitation with 100\% amplitude and -100\% amplitude respectively. Amplitude of displacement is inverted when excitation amplitude is inverted.%
}
\end{center}
\end{figure}

Figure \ref{Comparaison}-(A-B) illustrates Z-component map 6 ms after excitation, when excited in "monophasic" mode and in "biphasic" mode respectively. Average displacement in the region of interest is equal to 3.3 $\mu$m in the first case and 0.2 $\mu$m in the second case: only "monophasic" mode is able to induce observable shear waves.

Figure \ref{Comparaison}-(C-D) illustrates Z-component map 6 ms after excitation in a 5\% salt medium and in a 2\% medium respectively (note that (E) and (A) are identical). Average displacement in the region of interest is equal to 3.3 $\mu$m in the first case and 1.4 $\mu$m in the second case: when electrical conductivity of the medium decreases, shear wave amplitude decreases roughly by a same factor.

Figure \ref{Comparaison}-(E-F) illustrates Z-component map 6 ms after excitation, when excited with a 100\%  and 50\% amplitude in the coil respectively (note that (C) and (A) are identical). Average displacement in the region of interest is equal to 3.3 $\mu$m in the first case and 1.3 $\mu$m in the second case: shear wave amplitude is roughly divided by two when excitation amplitude is halved (according to the device panel).

Figure \ref{Comparaison}-(G-H) illustrates Z-component map 6 ms after excitation, when excited with a 100\% and a -100\% amplitude in the coil respectively, in the 2\% salt medium (note that (G) and (D) are identical). Average displacement in the region of interest is equal to 1.3 $\mu$m in the first case and -1.4 $\mu$m in the second case: displacement amplitude is inverted when excitation is inverted.

\begin{figure}[h!]
\begin{center}
\includegraphics[width=0.98\columnwidth]{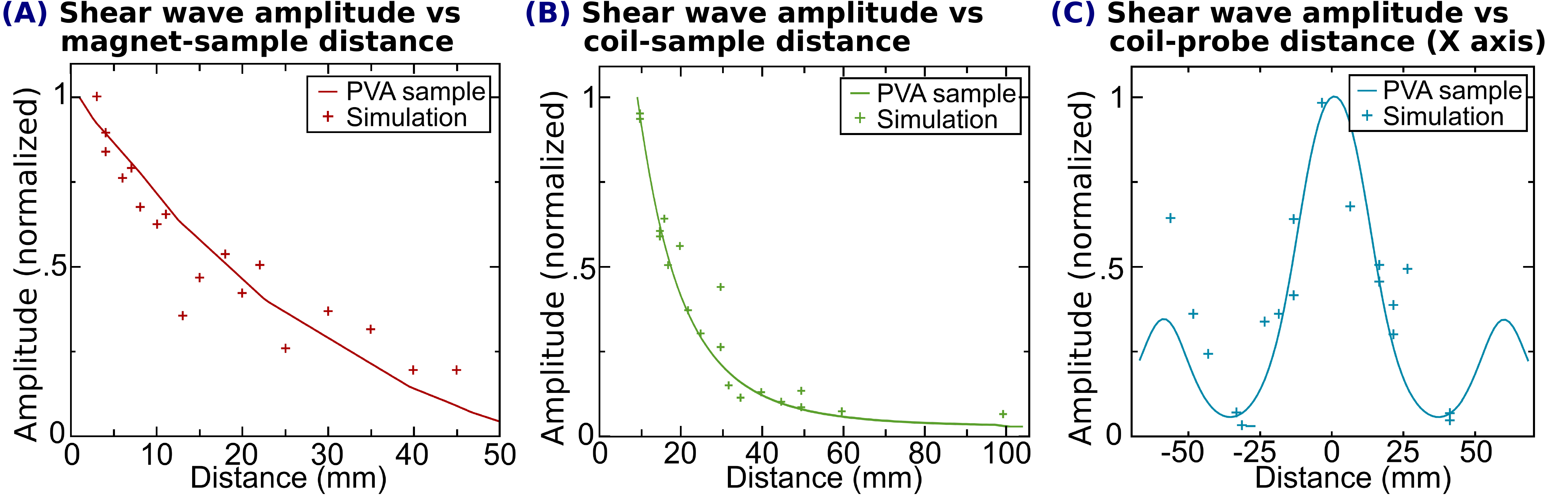}
\caption{\label{Figure5}(A) Normalized magnitude of shear waves versus distance along X from the magnet, experimental (red markers) and numerical (red line) results. (B) Normalized magnitude of shear waves versus distance along Z from the coil, experimental (green markers) and numerical (green line) results. (C) Normalized magnitude of shear waves versus distance along X between the center of the coil and the center of the probe, experimental (blue markers) and numerical (blue line) results.%
}
\end{center}
\end{figure}

The normalized amplitude of shear waves versus distance between the magnet and the PVA sample is illustrated in Figure \ref{Figure5}-(A); versus distance between the coil and the PVA sample along the Z axis in Figure \ref{Figure5}-(B); and versus distance between the center of the coil and the center of the ultrasound probe along X axis (0 being defined as the coil center aligned with the probe center) in Figure \ref{Figure5}-(C). Amplitude of shear waves was measured as the mean squared displacement between 15 and 25 mm of the coil inside the medium, an arbitrary location where shear waves had high amplitudes. Amplitudes were normalized by the maximum measured, respectively at a distance of 4 mm between the magnet and the sample, 10 mm between the coil and the medium and 0 mm between the center of coil and the center of probe.

We observed a decrease of the shear wave amplitude when the distance between the medium and the magnet increased, with an excellent agreement between experimental and numerical results. This was expected as the magnetic field decreased with distance. Similar observations can be made when the coil is drawn further from the sample. When we moved the coil along the X direction, we observed a strong maximum between two minima separated by 75 mm, corresponding to the length between the two centers of the TMS coil, which is also in agreement with the current density profile along this direction.

\section{Discussions}
\subsection{Practical application}
This study used an ultrasound device to image the sample and track shear waves, due to its high temporal resolution, availability and ease of use. However, for a clinical implementation such as brain elasticity imaging, MRI is more suited for tracking shear waves, as acoustic waves used in ultrasound imaging for shear wave tracking are attenuated by the skull. In a practical MRI implementation, no magnet would be necessary, and MRI-compatible coils should be used. As most clinical MRI scanners use 1.5 T or more magnetic fields, which is at least ten times higher than the one used in this study, displacement amplitude could thus be increased by a similar factor thus improving shear wave tracking.

Magnetic resonance elastography is usually employing continuous shear wave excitations. However, induction of a continuous electrical current in the TMS coil may interfere with MRI measurements, so repetitive triggered transient excitations may be used.

\subsection{Displacement amplitude}
In our numerical study, Lorentz force magnitude reached about 600 N.m$^{-3}$ for a 150 mT permanent magnetic field and a 7.5 S.m$^{-1}$ medium. The literatures provide numerous measurements of grey and white matter electrical conductivity. This parameter is however difficult to measure and thus presents a high variability. It indeed varied from 0.02 to 2 S.m$^{-1}$ depending on measurements \cite{19636081}. Using an average value of 0.2 S.m$^{-1}$, in a 1.5 T MRI system, the Lorentz force would reach a magnitude of about 160 N.m$^{-3}$. We can compare this value with the acoustic radiation force used for shear wave elastography. This force is calculated with the equation $f_{ARF} = 2 \alpha I / c$, where $\alpha$ is the attenuation in the medium, $I$ the ultrasound intensity and $c$ the speed of sound. Using Nightingale et al. parameters \cite{Nightingale_2001} ($\alpha$ = 0.4 Np.cm$^{-1}$, $I$ = 2.4 W.cm$^{-2}$, $c$ = 1540 m.s$^{-1}$),  $f_{ARF}$ is about 1200 N.m$^{-3}$, which led in their experimental study to displacements from 2.9 $\mu$m. The Lorentz force reported in the current study is about one order of magnitude smaller  but we could nevertheless observe displacements of 5 $\mu$m in the PVA phantom: it is indeed not only the amplitude, but also the shape and duration of excitation which contributes to the displacement magnitude.

Note that displacement reached an amplitude of 0.5 $\mu$m in the chicken sample. Electrical conductivity of muscle (longitudinal) is about 0.4 S.m$^{-1}$ and is expected to decrease notably after animal death. So although placed in saline water, the effective conductivity of the sample can be expected to be quite lower than that of saline (1.8 S.m$^{-1}$): this the probable explanation of the quite low amplitude displacement.

The excitation mode and duration also have an influence on the displacement amplitude. For example, the "biphasic" mode could not induce any observable displacement. This is probably due to the quick succession of positive and negative displacements with a mean at the noise level amplitude.

Finally, one could notice in the numerical study that displacements were slightly higher than the experimental values in the phantom. Various factors like viscosity and border effects, which were not included in our model, could explain this difference. Moreover, there were uncertainties about the electrical current amplitude and shape in the coil, as constructor values were used, and about the electrical conductivity of the medium, as this parameter is not entirely determined by the concentration of NaCl.

\subsection{Source localization}
In reported experiments, the shear wave source was 3 to 4 cm wide. With currently existing TMS coil geometries, it could hardly be lower than 1 cm. While this is higher than acoustic radiation force using a single point focus (1-2 mm), this last technique is hardly applicable in the brain because of the skull, as mentioned earlier. Compared to current magnetic resonance elastography methods using an external shaker, the shear wave source is far more localized. For whole brain elasticity measurements, having a source spreading on a few cm should not be a problem. For localized measurements with the proposed method, the shear wave source would be placed close to the region of interest, but not inside.

\subsection{Safety of the method}
Regarding the safety of the method, strong magnetic fields in MRI systems are considered biologically harmless \cite{schenck2000safety}. About the potential harmful effects of the electrical current induced by the TMS coil, safety guidelines based on clinical reports have also been provided for the TMS technique \cite{rossi2009safety}. These guidelines have been respected in the current study: the excitation amplitude stayed within the manufacturer's limits with less than one activation per second (no harmful effect is expected with this configuration). Main concerns would be with repetitive bursts of short repetition periods, as it could increase local temperature in the brain and potentially be detrimental for the instrument. The combination of electrical current induced by the TMS coil with strong magnetic fields can produce displacements in the range of a few tenths of micrometers at most in biological tissues, but no harmful effects have been reported so far with shear waves of this amplitude  \cite{skurczynski2009evaluation}, \cite{ehman2008vibration}. Precautions linked to the proposed technique are consequently the same as those for TMS and MRI -- mainly the absence of ferromagnetic materials in the body.

\section{Acknowledgments}

Part of the research has been founded by the Fondation pour la Recherche M\'edicale, the Natural Sciences and Engineering Research Council of Canada, the Fonds de Recherche du Qu\'ebec en Sant\'e and the Fondation de l'Association des Radiologistes du Qu\'ebec. The authors would like to thank MagVenture and Dr Paul Lesp\'erance for the loan of the TMS devices.

\bibliographystyle{plain}
\bibliography{biblio.bib}

\begin{thebibliography}{10}

\bibitem{aki1980quantitative}
Keiiti Aki and Paul~G Richards.
\newblock {\em {Quantitative Seismology}}.
\newblock Freeman San Francisco, 1980.

\bibitem{basford2005lorentz}
Alexandra~T Basford, Jeffrey~R Basford, Jennifer Kugel, and Richard~L Ehman.
\newblock {Lorentz-force-induced motion in conductive media}.
\newblock {\em Magnetic resonance imaging}, 23(5):647--651, 2005.

\bibitem{bercoff2004supersonic}
J{\'e}r{\'e}my Bercoff, Micka{\"e}l Tanter, and Mathias Fink.
\newblock {Supersonic shear imaging: a new technique for soft tissue elasticity
  mapping}.
\newblock {\em IEEE Transactions on Ultrasonics, Ferroelectrics and Frequency
  Control}, 51(4):396--409, 2004.

\bibitem{berg2012shear}
Wendie~A Berg, David~O Cosgrove, Caroline~J Dor{\'e}, Fritz~KW Sch{\"a}fer,
  William~E Svensson, Regina~J Hooley, Ralf Ohlinger, Ellen~B Mendelson,
  Catherine Balu-Maestro, Martina Locatelli, et~al.
\newblock {Shear-wave elastography improves the specificity of breast US: the
  BE1 multinational study of 939 masses}.
\newblock {\em Radiology}, 262(2):435--449, 2012.

\bibitem{bohning1997mapping}
D~E Bohning, A~P Pecheny, C~M Epstein, A~M Speer, D~J Vincent, W~Dannels, and
  M~S George.
\newblock {Mapping transcranial magnetic stimulation (TMS) fields in vivo with
  MRI}.
\newblock {\em Neuroreport}, 8(11):2535--2538, 1997.

\bibitem{Braun_2003}
Juergen Braun, Karl Braun, and Ingolf Sack.
\newblock {Electromagnetic actuator for generating variably oriented shear
  waves in {MR} elastography}.
\newblock {\em Magnetic Resonance in Medicine}, 50(1):220--222, jun 2003.

\bibitem{cobbold2007foundations}
Richard~SC Cobbold.
\newblock {\em {Foundations of Biomedical Ultrasound}}.
\newblock Oxford University Press, USA, 2007.

\bibitem{cochlin2002elastography}
D~Ll Cochlin, RH~Ganatra, and DFR Griffiths.
\newblock {Elastography in the detection of prostatic cancer}.
\newblock {\em Clinical Radiology}, 57(11):1014--1020, 2002.

\bibitem{devlin2003semantic}
J~Devlin, P~Matthews, and M~Rushworth.
\newblock {Semantic processing in the left inferior prefrontal cortex: a
  combined functional magnetic resonance imaging and transcranial magnetic
  stimulation study}.
\newblock {\em Journal of Cognitive Neuroscience}, 15(1):71--84, 2003.

\bibitem{ehman2008vibration}
EC~Ehman, PJ~Rossman, SA~Kruse, AV~Sahakian, and KJ~Glaser.
\newblock {Vibration safety limits for magnetic resonance elastography}.
\newblock {\em Physics in Medicine and Biology}, 53(4):925, 2008.

\bibitem{fromageau2007estimation}
J{\'e}r{\'e}mie Fromageau, Jean-Luc Gennisson, C{\'e}dric Schmitt, Roch~L
  Maurice, Rosaire Mongrain, Guy Cloutier, et~al.
\newblock {Estimation of polyvinyl alcohol cryogel mechanical properties with
  four ultrasound elastography methods and comparison with gold standard
  testings}.
\newblock {\em IEEE Transactions on Ultrasonics, Ferroelectrics, and Frequency
  Control}, 54(3):498--509, 2007.

\bibitem{19636081}
C~Gabriel, A~Peyman, and EH~Grant.
\newblock {Electrical conductivity of tissue at frequencies below 1 MHz.}
\newblock {\em Phys Med Biol}, 54:4863--78, Aug 2009.

\bibitem{gallot2011passive}
Thomas Gallot, Stefan Catheline, Philippe Roux, Javier Brum, Nicolas Benech,
  and Carlos Negreira.
\newblock {Passive elastography: shear-wave tomography from physiological-noise
  correlation in soft tissues}.
\newblock {\em IEEE Transactions on Ultrasonics, Ferroelectrics, and Frequency
  Control}, 58(6):1122--1126, 2011.

\bibitem{garcia2013stolt}
Damien Garcia, LL~Tarnec, St{\'e}phan Muth, Emmanuel Montagnon, Jonathan
  Por{\'e}e, and Guy Cloutier.
\newblock {Stolt's fk migration for plane wave ultrasound imaging}.
\newblock {\em IEEE Transactions on Ultrasonics, Ferroelectrics, and Frequency
  Control}, 60(9):1853--1867, 2013.

\bibitem{Grandori_1991}
F.~Grandori and P.~Ravazzani.
\newblock {Magnetic stimulation of the motor cortex-theoretical
  considerations}.
\newblock {\em {IEEE} Transactions on Biomedical Engineering}, 38(2):180--191,
  1991.

\bibitem{grandori1991magnetic}
Ferdinand Grandori and Paolo Ravazzani.
\newblock {Magnetic stimulation of the motor cortex-theoretical
  considerations}.
\newblock {\em IEEE Transactions on Biomedical Engineering}, 38(2):180--191,
  1991.

\bibitem{grasland2014elastoEMarticle}
Pol Grasland-Mongrain, R{\'e}mi Souchon, Ali Zorgani, Florian Cartellier,
  Jean-Yves Chapelon, Cyril Lafon, and Stefan Catheline.
\newblock {Imaging of shear waves induced by Lorentz force in soft tissues}.
\newblock {\em Physical Review Letters}, 113(3):038101, 2014.

\bibitem{hallett2000transcranial}
Mark Hallett.
\newblock {Transcranial magnetic stimulation and the human brain}.
\newblock {\em Nature}, 406(6792):147--150, 2000.

\bibitem{23008140}
S~Hirsch, D~Klatt, F~Freimann, M~Scheel, J~Braun, and I~Sack.
\newblock {In vivo measurement of volumetric strain in the human brain induced
  by arterial pulsation and harmonic waves.}
\newblock {\em Magn Reson Med}, 70:671--83, Sep 2013.

\bibitem{ilmoniemi1999transcranial}
Risto~J Ilmoniemi, Jarmo Ruohonen, and Jari Karhu.
\newblock {Transcranial magnetic stimulation-a new tool for functional imaging
  of the brain}.
\newblock {\em Critical Reviews in Biomedical Engineering}, 27(3-5):241--284,
  1999.

\bibitem{jackson1998classical}
John~David Jackson.
\newblock {\em {Classical Electrodynamics}}.
\newblock John Wiley and Sons, 3d edition, 1998.

\bibitem{kruse2008magnetic}
Scott~A Kruse, Gregory~H Rose, Kevin~J Glaser, Armando Manduca, Joel~P Felmlee,
  Clifford~R Jack~Jr, and Richard~L Ehman.
\newblock {Magnetic resonance elastography of the brain}.
\newblock {\em Neuroimage}, 39(1):231--237, 2008.

\bibitem{20833495}
P~Latta, ML~Gruwel, P~Debergue, B~Matwiy, UN~Sboto-Frankenstein, and B~Tomanek.
\newblock {Convertible pneumatic actuator for magnetic resonance elastography
  of the brain.}
\newblock {\em Magn Reson Imaging}, 29:147--52, Jan 2011.

\bibitem{mariappan2010magnetic}
Yogesh~K Mariappan, Kevin~J Glaser, and Richard~L Ehman.
\newblock {Magnetic resonance elastography: a review}.
\newblock {\em Clinical Anatomy}, 23(5):497--511, 2010.

\bibitem{FEMM}
D.C. Meeker.
\newblock {Finite Element Method Magnetics, Version 4.2, http://www.femm.info},
  2006.

\bibitem{montagnon2012real}
Emmanuel Montagnon, Sami Hissoiny, Philippe Despr{\'e}s, and Guy Cloutier.
\newblock {Real-time processing in dynamic ultrasound elastography: A GPU-based
  implementation using CUDA}.
\newblock In {\em 11th International Conference on Information Science, Signal
  Processing and their Applications (ISSPA)}, pages 472--477. IEEE, 2012.

\bibitem{murphy2011decreased}
Matthew Murphy, John Huston, Clifford Jack, Kevin Glaser, Armando Manduca, Joel
  Felmlee, and Richard Ehman.
\newblock {Decreased brain stiffness in Alzheimer's disease determined by
  magnetic resonance elastography}.
\newblock {\em Journal of Magnetic Resonance Imaging}, 34(3):494--498, 2011.

\bibitem{muthupillai1995magnetic}
R~Muthupillai, DJ~Lomas, PJ~Rossman, JF~Greenleaf, A~Manduca, and RL~Ehman.
\newblock {Magnetic resonance elastography by direct visualization of
  propagating acoustic strain waves}.
\newblock {\em Science}, 269(5232):1854--1857, 1995.

\bibitem{nightingale2002acoustic}
Kathryn Nightingale, Mary~Scott Soo, Roger Nightingale, and Gregg Trahey.
\newblock {Acoustic radiation force impulse imaging: in vivo demonstration of
  clinical feasibility}.
\newblock {\em Ultrasound in Medicine and Biology}, 28(2):227--235, 2002.

\bibitem{Nightingale_2001}
Kathryn~R. Nightingale, Mark~L. Palmeri, Roger~W. Nightingale, and Gregg~E.
  Trahey.
\newblock {On the feasibility of remote palpation using acoustic radiation
  force}.
\newblock {\em The Journal of the Acoustical Society of America}, 110(1):625,
  2001.

\bibitem{rossi2009safety}
Simone Rossi, Mark Hallett, Paolo~M Rossini, and Alvaro Pascual-Leone.
\newblock {Safety, ethical considerations, and application guidelines for the
  use of transcranial magnetic stimulation in clinical practice and research}.
\newblock {\em Clinical Neurophysiology}, 120(12):2008--2039, 2009.

\bibitem{sakkas2006repetitive}
Pavlos Sakkas, Constantin Psarros, George~N Papadimitriou, Christos~G
  Theleritis, and Constantin~R Soldatos.
\newblock {Repetitive transcranial magnetic stimulation (rTMS) in a patient
  suffering from comorbid depression and panic disorder following a myocardial
  infarction}.
\newblock {\em Progress in Neuro-Psychopharmacology and Biological Psychiatry},
  30(5):960--962, 2006.

\bibitem{sandrin2003transient}
Laurent Sandrin, Bertrand Fourquet, Jean-Michel Hasquenoph, Sylvain Yon,
  C{\'e}line Fournier, Fr{\'e}d{\'e}ric Mal, Christos Christidis, Marianne
  Ziol, Bruno Poulet, Farad Kazemi, Michel Beaugrand, and Robert Palau.
\newblock {Transient elastography: a new noninvasive method for assessment of
  hepatic fibrosis}.
\newblock {\em Ultrasound in Medicine and Biology}, 29(12):1705--1713, 2003.

\bibitem{sarvazyan1998shear}
Armen~P Sarvazyan, Oleg~V Rudenko, Scott~D Swanson, J~Brian Fowlkes, and
  Stanislav~Y Emelianov.
\newblock {Shear wave elasticity imaging: a new ultrasonic technology of
  medical diagnostics}.
\newblock {\em Ultrasound in Medicine and Biology}, 24(9):1419--1435, 1998.

\bibitem{schenck2000safety}
John~F Schenck.
\newblock {Safety of strong, static magnetic fields}.
\newblock {\em Journal of Magnetic Resonance Imaging}, 12(1):2--19, 2000.

\bibitem{schmitt2010ultrasound}
C{\'e}dric Schmitt, Anis~Hadj Henni, and Guy Cloutier.
\newblock {Ultrasound dynamic micro-elastography applied to the viscoelastic
  characterization of soft tissues and arterial walls}.
\newblock {\em Ultrasound in medicine \& biology}, 36(9):1492--1503, 2010.

\bibitem{skurczynski2009evaluation}
MJ~Skurczynski, FA~Duck, JA~Shipley, JC~Bamber, and D~Melodelima.
\newblock {Evaluation of experimental methods for assessing safety for
  ultrasound radiation force elastography}.
\newblock {\em The British journal of radiology}, 2014.

\bibitem{taylor2004reassessment}
Zeike Taylor and Karol Miller.
\newblock {Reassessment of brain elasticity for analysis of biomechanisms of
  hydrocephalus}.
\newblock {\em Journal of Biomechanics}, 37(8):1263--1269, 2004.

\bibitem{11548931}
JB~Weaver, Houten~EE Van, MI~Miga, FE~Kennedy, and KD~Paulsen.
\newblock {Magnetic resonance elastography using 3D gradient echo measurements
  of steady-state motion.}
\newblock {\em Med Phys}, 28:1620--8, Aug 2001.

\bibitem{Weaver_2012}
John~B Weaver, Adam~J Pattison, Matthew~D McGarry, Irina~M Perreard, Jessica~G
  Swienckowski, Clifford~J Eskey, S~Scott Lollis, and Keith~D Paulsen.
\newblock {Brain mechanical property measurement using {MRE} with intrinsic
  activation}.
\newblock {\em Physics in Medicine and Biology}, 57(22):7275--7287, oct 2012.

\bibitem{wuerfel2010mr}
Jens Wuerfel, Friedemann Paul, Bernd Beierbach, Uwe Hamhaber, Dieter Klatt,
  Sebastian Papazoglou, Frauke Zipp, Peter Martus, J{\"u}rgen Braun, and Ingolf
  Sack.
\newblock {MR-elastography reveals degradation of tissue integrity in multiple
  sclerosis}.
\newblock {\em Neuroimage}, 49(3):2520--2525, 2010.

\bibitem{Zorgani_2015}
Ali Zorgani, R{\'{e}}mi Souchon, Au-Hoang Dinh, Jean-Yves Chapelon, Jean-Michel
  M{\'{e}}nager, Samir Lounis, Olivier Rouvi{\`{e}}re, and Stefan Catheline.
\newblock {Brain palpation from physiological vibrations using {MRI}}.
\newblock {\em Proceedings of the National Academy of Sciences},
  112(42):12917--12921, oct 2015.

\end{thebibliography}

\end{document}